\documentclass{PoS}

\title{The CMS muon system in Run2: preparation, status and first results}

\ShortTitle{The CMS muon system in Run2: preparation, status and first results}

\author{\speaker{Giovanni Abbiendi}%
        \thanks{on behalf of the CMS collaboration}\\
       INFN - Sezione di Bologna\\
       E-mail: \email{giovanni.abbiendi@cern.ch}}

\abstract{
The CMS muon system has played a key role for many physics results obtained from the LHC Run-1 data. 
During the Long Shutdown (2013-2014) significant upgrades have been carried out on the muon detectors
and on the L1 muon trigger.  
In parallel the algorithms for muon reconstruction and identification have been improved for both the 
High-Level Trigger and the offline reconstruction. 
Results of studies performed on data and Monte Carlo simulations will be presented, 
with focus on the improvements aiming to ensure an excellent performance in conditions of multiplicity
of pileup events and bunch spacing expected during the high-luminosity phase of Run-2. 
The early muon performance results from LHC Run-2 will be shown. 
}

\FullConference{The European Physical Society Conference on High Energy Physics\\
		 22-29 July 2015\\
		 Vienna, Austria}

\usepackage{subfigure}

\begin{document}
\hyphenation{cal-or-i-me-ter}

\def\pt{p_T}
\def\Zo{\ensuremath{\mathrm {Z}}}
\def\W{\ensuremath{\mathrm{W}}}%
\def\mm{\mu^+\mu^-}%
\def\ppZ{\pp\to\Zo + X}%
\def\ppbarW{\ppbar\to\W + X}%
\def\ppbarZ{\ppbar\to\Zo + X}%
\def\ppW{\pp\to\W + X}%
\def\ppZGmm{\pp\to\Zo(\gamma^*) + X \to \mm + X}%
\def\ppZmm{\pp\to\Zo + X \to \mm + X}%
\def\ppWmn{\pp\to\W + X \to \mu\nu + X}%
\def\ppWtn{\pp\to\W + X \to \tau\nu + X}%
\def\Zll{\Zo\to \ell\ell}%
\def\Zmm{\Zo\to\mm}%
\def\Umm{\Upsilon \to\mm}%
\def\Wln{\W\to\ell\nu}%
\def\Wmn{\W\to\mu\nu}%
\def\Wpmn{\W^+\to\mu^+\nu_\mu}%
\def\Wmmn{\W^-\to\mu^-\bar{\nu}_\mu}%
\newcommand{\jpsi}{\ensuremath{\mathrm{J}/\!\psi}}
\def\jpsimm{\jpsi\to\mm}%
\newcommand{\IRelComb} {I^{\text{rel}}_{\text{comb}}}%
\newcommand{\ITRK}     {I_{\text{trk}}}%
\newcommand{\IECAL}    {I_{\text{ECAL}}}%
\newcommand{\IHCAL}    {I_{\text{HCAL}}}%
\newcommand{\IRelTrk} {I^{\text{rel}}_{\text{trk}}}%
\newcommand{\IPF} {I^{\text{rel}}_{\textrm{PF}}}%

\section{Muon detectors and upgrades during the Long Shutdown 1}
\begin{figure}[htbp]
{\centering
\includegraphics[width=0.65\textwidth]{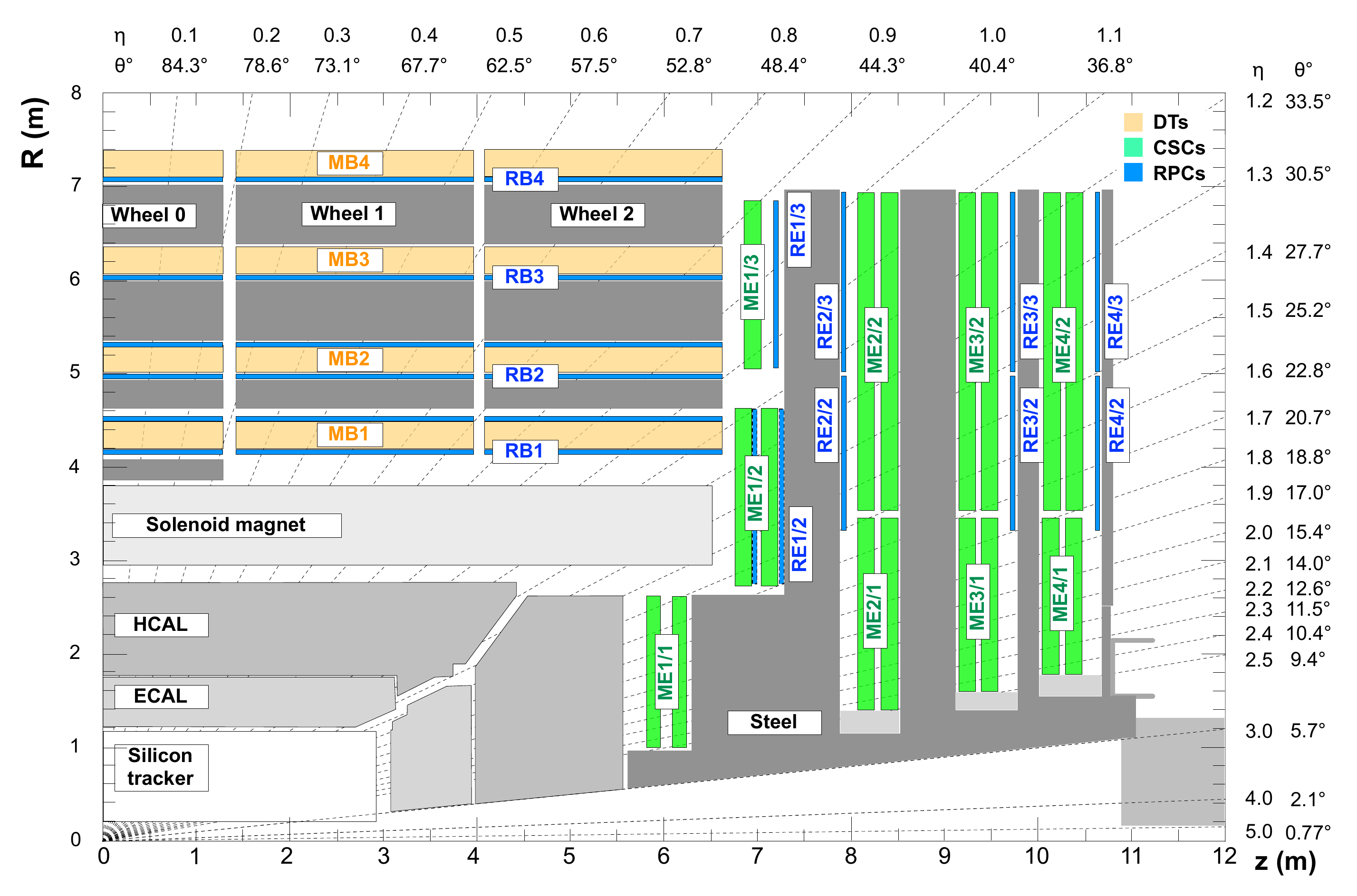}
\caption{\label{fig:CMSquadrant}
$R$--$z$ cross section of a quadrant of the CMS detector. The interaction point is at the lower left corner.}
}
\end{figure}

Muon detection is a powerful tool for recognizing signatures of interesting physics processes 
over the high background rates at the Large Hadron Collider (LHC).
A schematic view of the CMS detector \cite{cms} is shown in Fig.\ref{fig:CMSquadrant}.
Muon reconstruction is performed using the silicon tracker
at the centre of the detector immersed in a 3.8~T solenoidal magnetic
field, and with four stations of gas-ionization
muon detectors 
installed outside the solenoid and sandwiched
between the layers of the steel return yoke.
The inner tracker is composed of pixel and strip silicon detectors
and measures charged-particle trajectories in the pseudorapidity range
$|\eta|<2.5$.
The muon system consists of detectors with three different technologies \cite{muondet}.
Drift tube (DT) chambers in the barrel region ($|\eta|<1.2$), cathode strip
chambers (CSC) in the endcap regions ($0.9 <|\eta|< 2.4$), and resistive
plate chambers (RPC) covering the range of $|\eta|<1.8$.

Major upgrades of the muon detectors have been carried out in 2013-14 during the Long Shutdown 1 (LS1) \cite{guiducci}.
The outermost rings of CSC and RPC chambers in the fourth station of both endcaps have been
completed (ME4/2, RE4/2 and RE4/3. Previously only few CSC chambers had been installed and operated during Run1). 
The new rings cover the angular region $1.2 < \eta < 1.8$. 
With four stations the system redundancy is increased and the minimal requirements to build a L1 trigger track are met more often.
This allows to tighten the trigger quality cuts, keeping reasonably low thresholds at high luminosity.
To protect against environmental noise the endcaps have been shielded with outer yoke disks on both sides.
Another CSC upgrade affected the innermost rings in the first station (ME1/1A), at  $2.1 < \eta < 2.4$. 
Here the readout electronics has been refurbished to make use of the full detector granularity 
(in Run1 the readout strips were ganged in groups of three),  
allowing an improved performance of both trigger and offline reconstruction.
The DT chambers underwent a relocation of the trigger electronics from the experimental cavern to the service cavern, 
which gives more freedom to intervene at running time, in addition to improvements to the L1 Trigger, 
which will be completely upgraded in 2016.

\section{Run2 scenarios}
The LHC operation in Run2 is expected to be challenging for CMS,
due to the simultaneous increase of centre-of-mass energy (from 8~TeV to 13~TeV),
luminosity (peak from $\sim$7$\times 10^{33}$~cm$^{-2}$s$^{-1}$ to $\sim$1.5$\times 10^{34}$~cm$^{-2}$s$^{-1}$), 
and collision frequency (bunch spacing from 50~ns to 25~ns).
The changes sum up to give about a factor four larger physics rates in comparison to Run1.
In addition the higher luminosity implies an increased pileup of independent proton collisions at each bunch crossing.
The highest average pileup rate reached at the end of Run1 was about 30,
while the expectation for Run2 at design luminosity is about 40.
The reduced bunch spacing increases effects of pileup from out-of-time collisions.
To cope with the LHC schedule for 2015, three scenarios were defined corresponding to suitable trigger menus for the
evolving LHC conditions.
Owing to the improvements during the LS1 the initial thresholds for unprescaled L1 muon triggers can be kept 
at the same transverse momentum values as in the highest luminosity configuration of the 2012 trigger menu. 
The single-muon trigger has kept the L1 threshold $p_T = $16~GeV, which will be eventually raised to 20~GeV
at the maximum luminosity. 

\section{Reconstruction improvements}
The standard CMS muon reconstruction \cite{muonreco} starts from the local reconstruction on single detector elements, as
strip and pixel hits in the inner tracker, muon hits and/or segments on the muon chambers. 
Tracks are then reconstructed independently in the inner tracker ({\em tracker track}) and in the muon system ({\em standalone track}).
For each standalone-muon track a matching tracker track is found by comparing parameters of the two tracks propagated onto a common surface.
A {\em global-muon track} is fitted combining hits from the tracker track and standalone-muon track, using the Kalman-filter technique.
This {\em outside-in} algorithm is accompanied by an {\em inside-out} identification algorithm, 
where candidate tracker tracks are extrapolated to the muon system taking into account the magnetic field, the average expected energy losses, 
and multiple Coulomb scattering in the detector material. 
If at least one muon segment matches the extrapolated track, the corresponding tracker track qualifies as a {\em Tracker Muon}.

\subsection{High-Level Trigger}
\begin{figure}[htbp]
\begin{center}
 \subfigure[]{\includegraphics[width=0.5\textwidth]{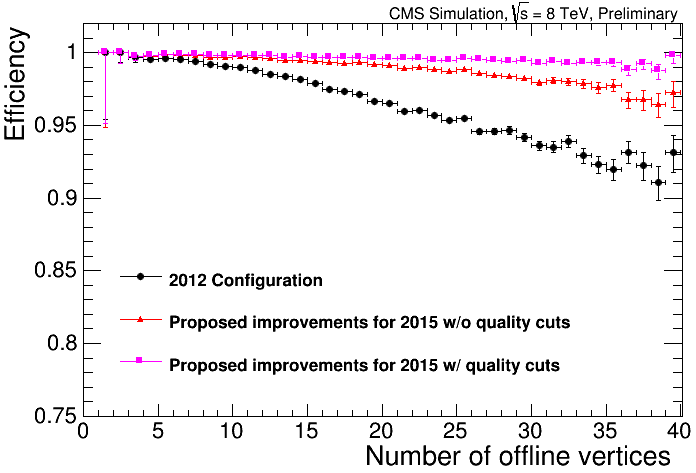}}
 \subfigure[]{\includegraphics[width=0.4\textwidth]{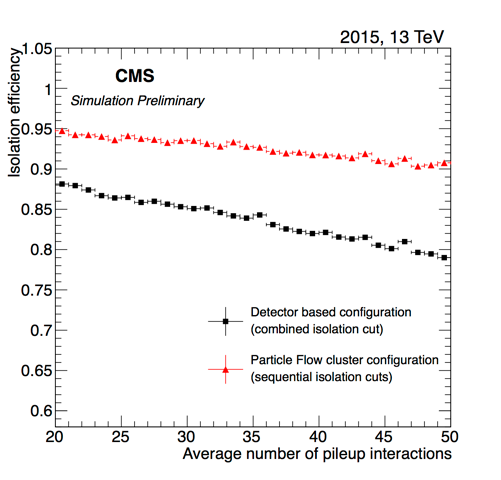}}
\caption{\label{fig:HLT}
(a) Muon HLT efficiency for the global reconstruction step as a function of the number of reconstructed vertices, 
from a MC simulation of $W \to \mu\nu$ events. 
(b) Muon HLT isolation efficiency as a function of the average number of pileup interactions for two working points giving a similar rate reduction.
}
\end{center}
\end{figure}

The reconstruction proceeds through the same basic steps in the online HLT and in the offline reconstruction, though the online algorithms
have to be faster and hence are less sophisticated \cite{perrotta}.
The HLT global muon reconstruction is designed to be fast and efficient for prompt muons. Its core is a three-fold algorithm 
for seeding and pattern recognition, with the first two branches starting from outside and going inward and 
the last one starting from inside and going outward. The three branches are run from the fastest to the slowest and the cascade 
is interrupted when a candidate muon track is reconstructed. 
Quality cuts may be applied, in particular to test the compatibility of the muon track with the beam spot 
(e.g. by a 1-2~mm cut on the transverse impact parameter).
A degraded performance had been observed for the HLT global muon reconstruction at the end of Run1,                  
causing a significant inefficiency of $5-10$\% for high pileup values.
In Run2 the pattern recognition has been improved by replacing a simple geometrical criterion with a $\chi^2$ matching to assign hits to the tracker track.
In addition the quality cuts have been moved to be applied as the last step of each branch, to avoid reconstructing fake tracks which 
would then be cut off at the end of the cascade. 
The results obtained from MC simulations are shown in Fig.~\ref{fig:HLT}(a).
With these changes the pileup-dependent inefficiency has been completely recovered.
The single-muon HLT thresholds will stay at $p_T = 45-50$~GeV for the inclusive non-isolated muon trigger.
Other significant improvements have been implemented in the inner tracking, which has become faster and better in Run2. 
This has also allowed to deploy Tracker Muon triggers with a similar {\em Inside-Out} algorithm as that used in the offline muon identification.

Single-muon triggers must keep under control the rate of background muons from QCD processes. Muon isolation is the key quantity for this. 
During Run1, muon isolation in HLT was calculated by summing up tracks and calorimetric energy deposits around the muon direction 
({\em detector-based isolation}). A better performance is obtained by clustering the calorimetric energy deposits according to the 
particle-flow event reconstruction \cite{PFalgo,PFmuons} and then
by the application of sequential cuts on the isolation in the electromagnetic calorimeter, hadronic calorimeter and inner tracker. 
Fig.~\ref{fig:HLT}(b) shows the efficiency of selecting $\gamma/Z \rightarrow \mm$ events as a function of the average number 
of pileup interactions, for a given rejection of the QCD backgrounds. 
As visible the efficiency is increased and the pileup dependence is reduced. 

These improvements among others allow to keep the same HLT trigger threshold as in 
Run1 for the isolated single-muon trigger, $p_T = 24$~GeV for $|\eta| < 2.1$, 
in the scenario corresponding to the maximum luminosity.
The extreme endcap regions ($2.1<|\eta|<2.4$) will require slightly higher thresholds.
Muon isolation has also been introduced in double muon triggers, where it can 
reduce significantly the rates with negligible losses.
In this way double muon triggers will keep $p_T$ thresholds at $17 - 8$~GeV as in Run1.
The overall quality of the Run2 trigger has improved significantly with respect to Run1: 
it has better efficiency and purity while the total expected rate does not exceed the design value.

\subsection{Offline reconstruction}
\begin{figure}[htbp]
\begin{center}
 \subfigure[]{\includegraphics[width=0.35\textwidth]{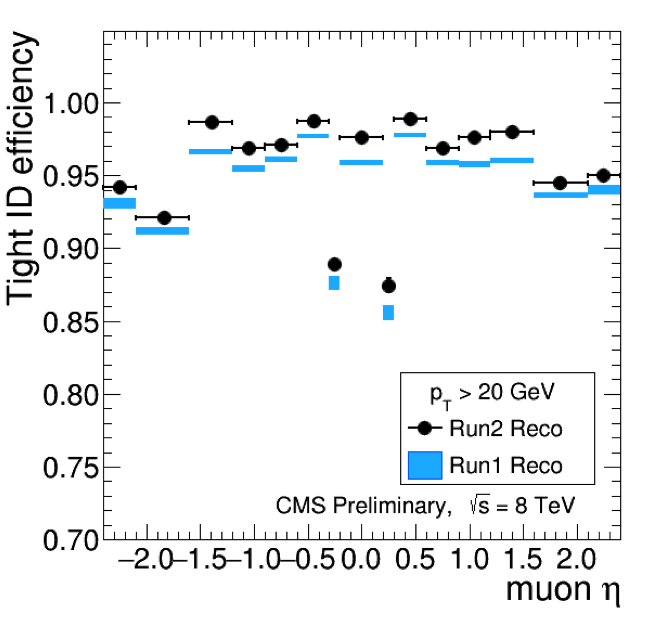}}
 \subfigure[]{\includegraphics[width=0.35\textwidth]{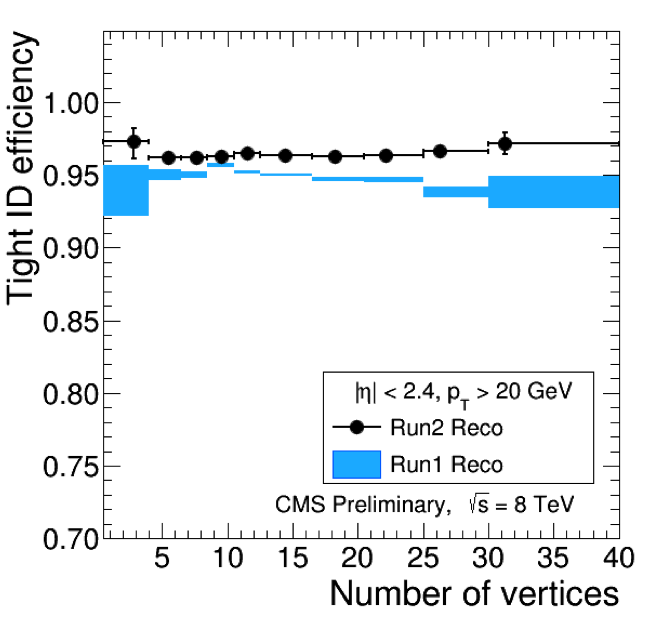}}
\caption{\label{fig:TightIDeff}
Tag-and-Probe efficiency for Tight Muon identification on Run1 data reconstructed with both the old and the new algorithm: 
(a) as a function of pseudorapidity for muon $p_T >$~20~GeV;
(b) as a function of the number of recostructed vertices for muon $p_T >$~20~GeV and $|\eta| < 2.4$.
}
\end{center}
\end{figure}

A small efficiency loss had been observed during Run1 for muon tracker tracks, increasing with pileup.
To cure this behaviour two muon-specific iterations have been implemented within the general tracking \cite{TRK-paper}: the first one outside-in, 
seeded from the muon system to recover missing tracker tracks; the second one inside-out, to reconstruct muon-tagged tracks with looser requirements, 
thus improving the hit collection efficiency. In addition to recovering the small inefficiency the new algorithm has also improved the muon track 
quality, in particular the number of hits per track. This is reflected in the identification efficiency of standard selections 
as the Tight Muon selection \cite{DPS-POG}, 
shown in Fig.~\ref{fig:TightIDeff}. An efficiency gain of about 2\% is observed.

\begin{figure}[htbp]
\begin{center}
 \subfigure[]{\includegraphics[width=0.35\textwidth]{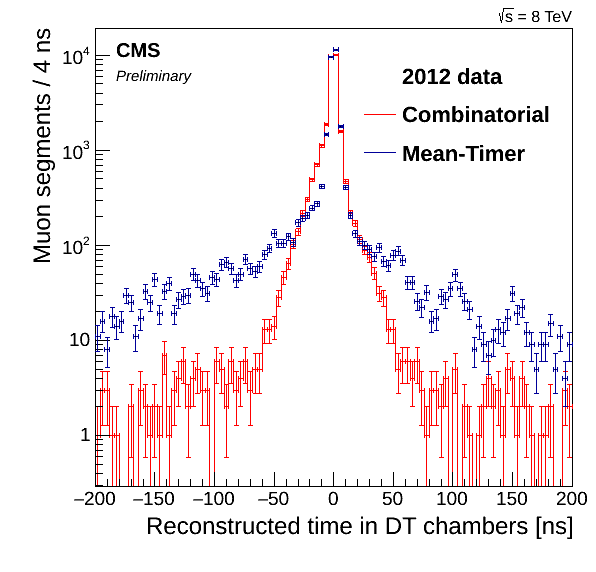}}
 \subfigure[]{\includegraphics[width=0.45\textwidth]{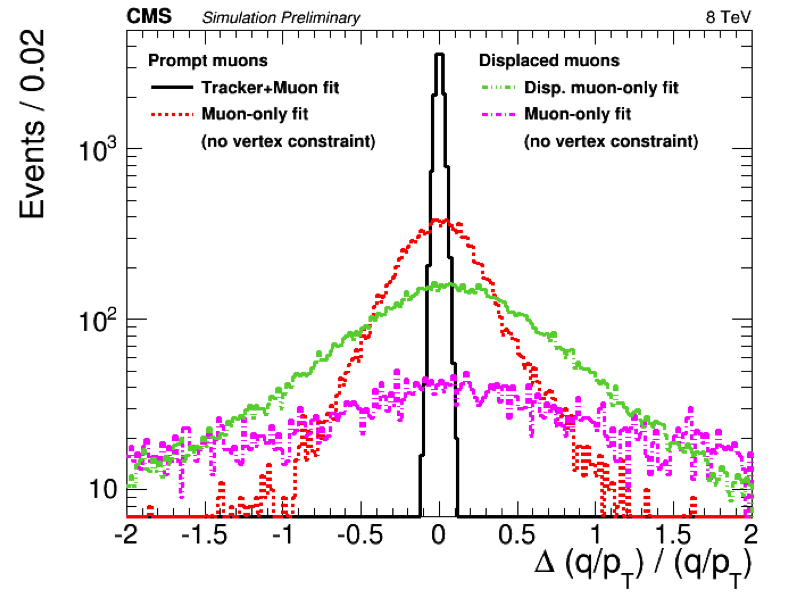}} \\
 \subfigure[]{\includegraphics[width=0.4\textwidth]{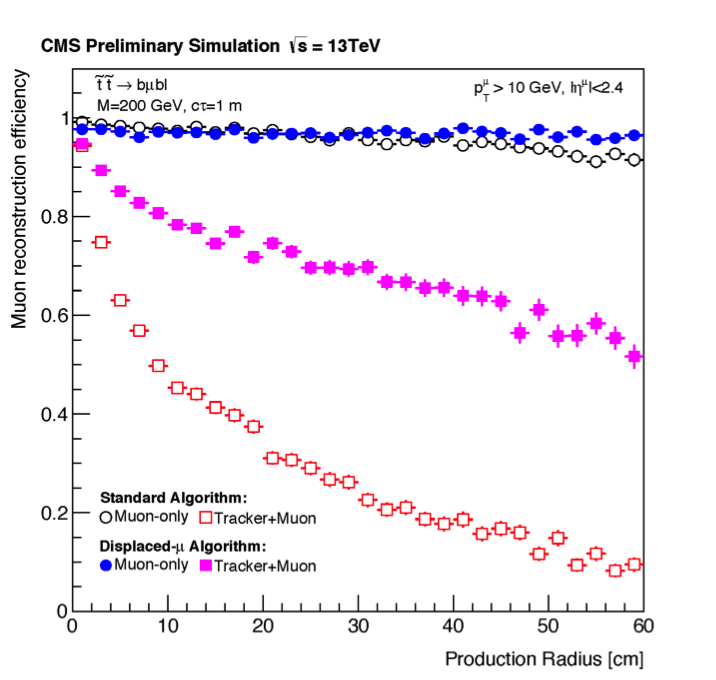}}
 \subfigure[]{\includegraphics[width=0.4\textwidth]{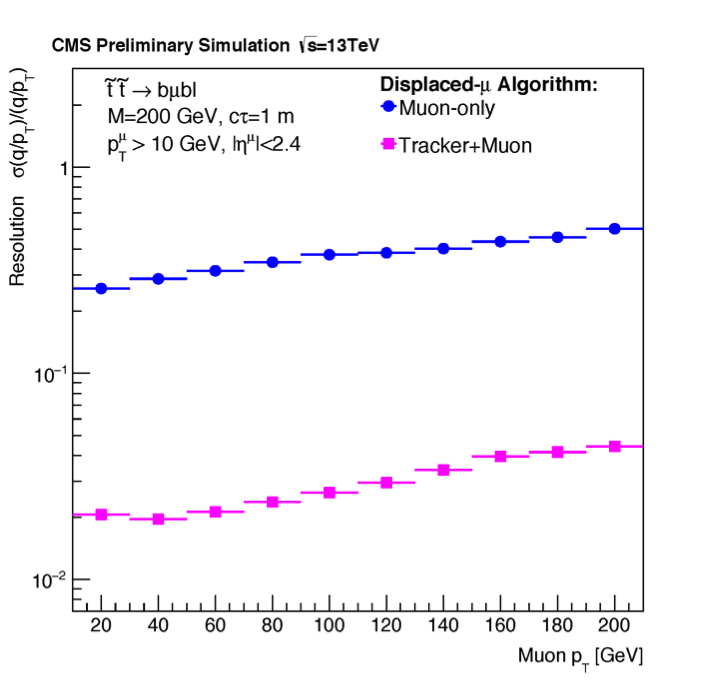}}
\caption{\label{fig:DTmuons} 
(a) Muon segment time in DT chambers for Run1 data; 
(b) $q/p_T$ relative residual for displaced muons from a simulated exotic decay 
with tipical decay length greater than 1~m and for prompt muons with similar momentum, for comparison;
(c) Muon reconstruction efficiency as a function of the production radius.
The standard muon-only and tracker+muon algorithms are compared to the new algorithms specific for displaced muons.
(d) $q/p_T$ resolution as a function of muon $p_T$. 
}
\end{center}
\end{figure}

Local reconstruction in the barrel muon chambers (DT) has also been improved for Run2.
During Run1 DT segments were formed by a combinatorial pattern recognition, 
assuming an in-time muon traversing a chamber and fitting the individual hits with a straight line.
Owing to the geometry of DT's, with half-cell staggered layers, and the almost constant drift velocity 
the {\em Mean-Timer} algorithm can be exploited \cite{DPS-DTDPG}.
The time, at which the particle crossing occurs, can be kept as a free parameter in the segment fit. 
This algorithm improves the $\delta$-ray rejection and the position resolution, while it provides in addition an excellent measurement 
of the segment time, with $\sim$2~ns resolution for in-time muons. It can be used to identify out-of-time muons, 
with time delays up to $\sim$10 bunch crossings, and improve their rejection. 
The performance has been tested on a sample of real data from Run1 processed with both the algorithms, as shown in Fig.~\ref{fig:DTmuons}(a). 
It is visible both a sharpening of the central peak, due to in-time muons, and an increased efficiency for out-of-time muons, 
where the 50~ns spacing between consecutive bunch crossings is clearly visible on the right side.

Improving the standalone muon performance is relevant for searches of exotic particles, like massive particles decaying to muons very far 
from the interaction point, possibly even delayed with respect to the time of the interaction. 
If muons are produced outside the inner-tracker volume, they can be detected only by muon chambers.
Moreover the inner tracker is efficient only for in-time particles. 
A new muon-only algorithm \cite{DPS-POG} has been developed for displaced (and/or delayed) tracks, 
with a specific seeding step similar to that used for the reconstruction of cosmic-ray muons, 
to avoid biases towards the interaction point. The current performance is shown in Fig.~\ref{fig:DTmuons}(b) in comparison 
to the standard muon-only algorithm, for decay muons from a simulated long-lived particle $X$ (with mass $M = 500$~GeV and $Q=2e$) 
with typical decay length greater than 1~m. 
The $p_T$ resolution is greatly improved for displaced muons, although it is still 
not as good as the one for the similar muon-only fit for prompt muons.
In Run2 this algorithm is applied also at trigger level, to select possible delayed muon signatures.

Another new algorithm \cite{DPS-POG} has been introduced for displaced in-time muons produced within the tracker volume, with the muon leaving hits 
both in the inner tracker and in the muon chambers. 
The standard tracker reconstruction is not optimized for displaced particles, therefore a specific tracking was implemented, 
seeded by the displaced standalone muons and not using any constraints to the interaction point in the pattern recognition nor in the track fit.
The reconstruction efficiency as a function of the production radius is significantly increased, as shown in Fig.~\ref{fig:DTmuons}(c) 
obtained from PYTHIA $\tilde t$-pair production followed by $\tilde t \rightarrow b + \mu$, with $M(\tilde t)=$200~GeV and $c\tau=$1~m 
and full detector simulation with Run2 conditions.
The final tracker+muon fit greatly improves the momentum resolution 
with respect to the muon-only fit, as shown in Fig.~\ref{fig:DTmuons}(d).

\section{Performance with the first 2015 data}
\begin{figure}[htbp]
\begin{center}
 \subfigure[]{\includegraphics[width=0.5\textwidth]{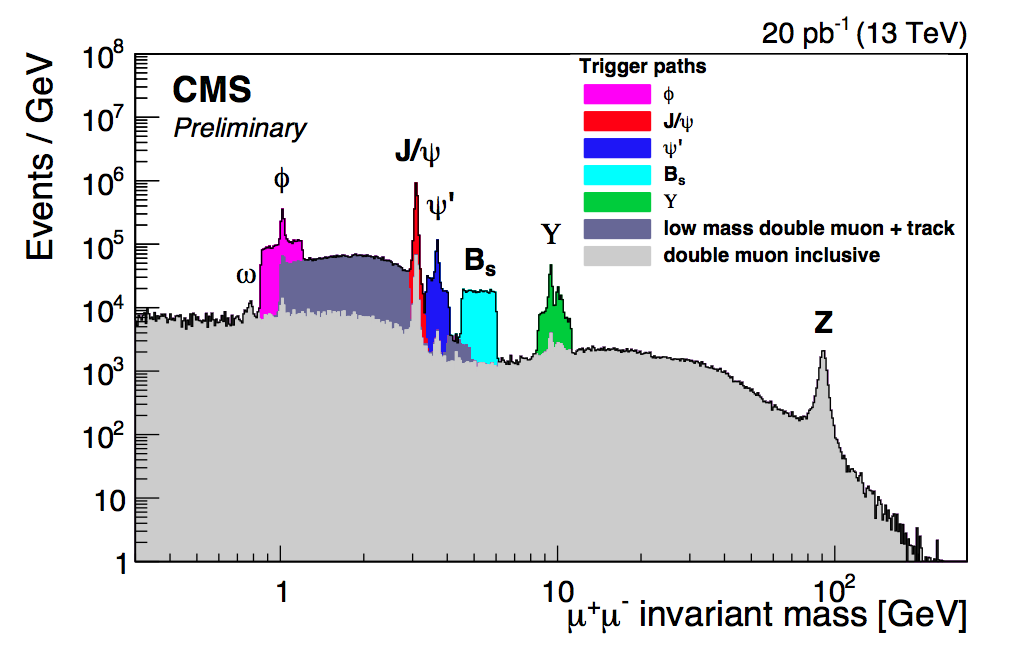}}
 \subfigure[]{\includegraphics[width=0.4\textwidth]{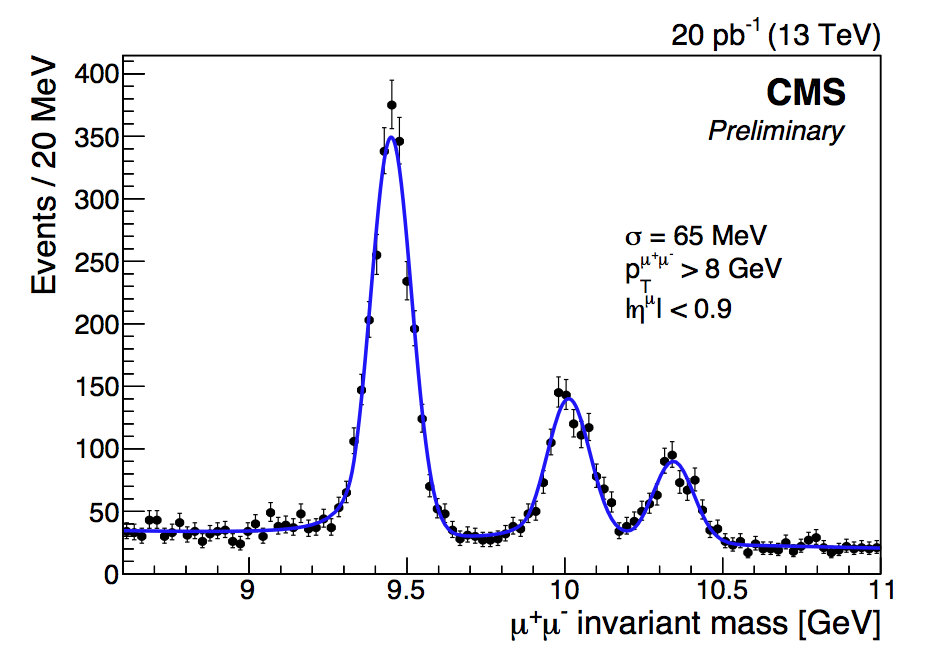}} \\
 \subfigure[]{\includegraphics[width=0.4\textwidth]{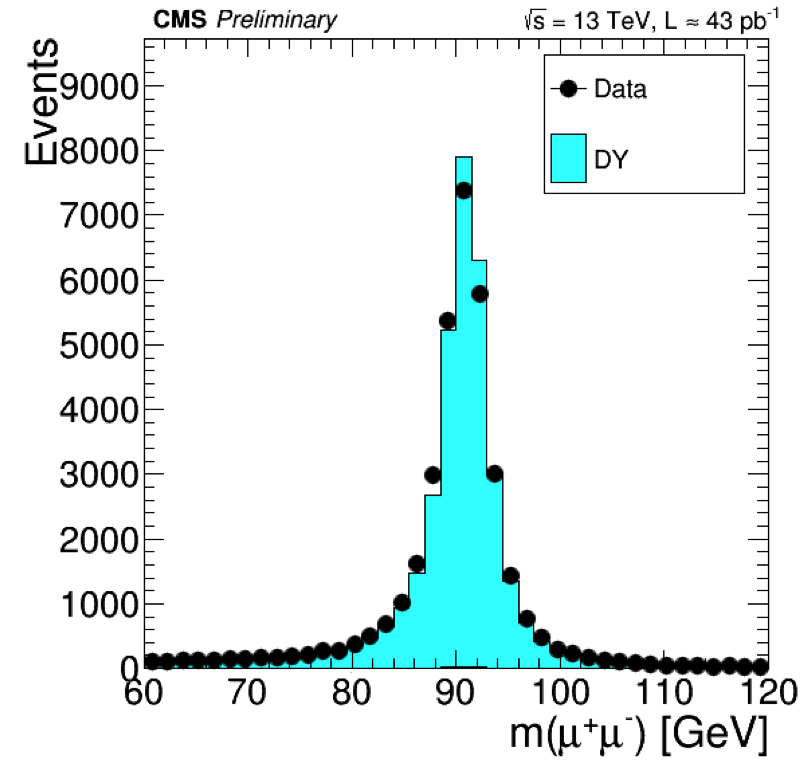}}
 \subfigure[]{\includegraphics[width=0.4\textwidth]{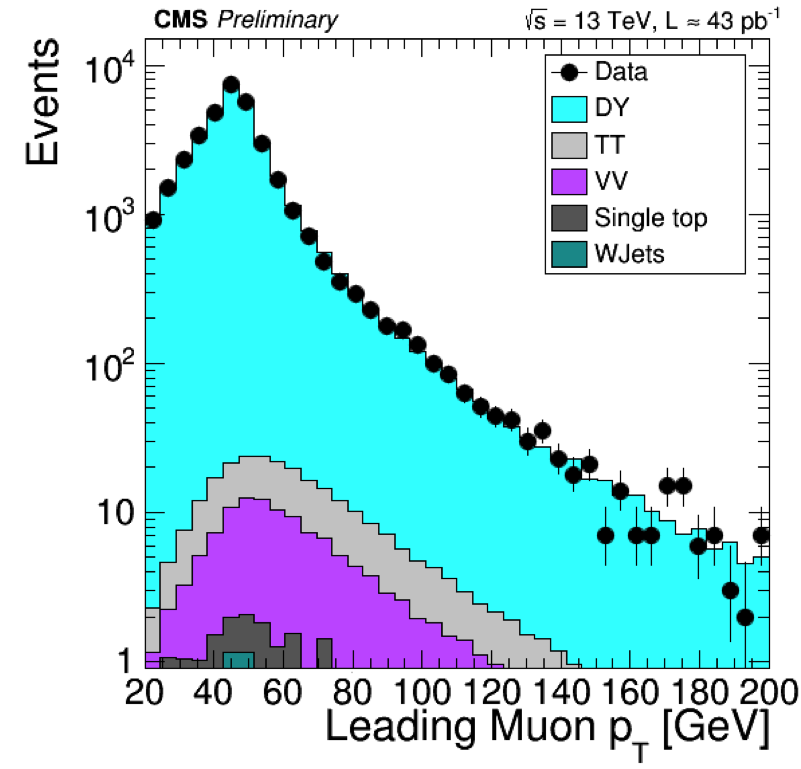}}
\caption{\label{fig:BPHandZ} 
(a) Dimuon mass spectrum from events collected with inclusive (gray) and specialized (colour) muon triggers in the first 2015 data;
(b) the $\Upsilon$ peaks after standard offline cuts with muons in the barrel region.
(c) The Z peak after standard selection cuts and (d) distribution of the leading muon $p_T$ from Z decay, both 
compared to Monte Carlo simulations normalized to the same number of events.
}
\end{center}
\end{figure}

The very first data from Run2 $pp$ collisions at 13~TeV have been collected in June 2015. 
Here a few early performance results are shown, obtained from $\sim 20 - 40$~pb$^{-1}$ \cite{DPS-POG, DPS-BPH}. 
Fig.~\ref{fig:BPHandZ}(a) shows the yield of reconstructed dimuon events as a function of the invariant mass. 
The events were selected by both inclusive and specialized dimuon triggers. 
The $\Upsilon$ mass region is shown with standard quality cuts applied for muons in the barrel region in Fig.~\ref{fig:BPHandZ}(b).
Though not exactly comparable, the performance is 
in line with that at start of Run1.
A clean signal from Z decays is observed in the first data (Fig.~\ref{fig:BPHandZ}-c), 
still with startup alignment conditions and no momentum calibration. 
The MC simulation is based on asymptotic conditions and is then optimistic, but in decent agreement with real data.
The $p_T$ distribution of the leading muon (Fig.~\ref{fig:BPHandZ}-d) is well described by the simulation.

\section{Conclusions}
Significant upgrades to the muon detectors and trigger (as well as to the inner tracker and other subdetectors) 
have been carried out during the Long Shutdown in 2013-14. 
CMS has been re-commissioned,
synchronized and calibrated with the first data, cosmics and collisions with magnet on and off.
The detector alignment will be improved by using more collision data.
Parallel improvements have been implemented to the high-level trigger and the offline recontruction to cope with the increased pileup 
and the LHC operation with 25~ns bunch spacing.
The first performance results are good.
Muon detection will retain its crucial role in exploring the exciting physics ahead of us.

\end{document}